\documentstyle[preprint,aps,psfig]{revtex}
\tighten
\draft
\widetext
\preprint{CLNS 99/1641}

\def\simgreat{\mathbin{\lower 3pt\hbox
   {$\rlap{\raise 5pt\hbox{$\char'076$}}\mathchar"7218$}}}

\def\bz{\beta_0}
\def\bo{\beta_1}
\def\b2{\beta_2}
\def\az{\alpha_0}
\def\ao{\alpha_1}
\def\a2{\alpha_2}
\def\Lb{\Lambda_b}
\def\that{\hat t}
\def\dthat{d\that}
\def\be{\begin{equation}}
\def\ee{\end{equation}}
\def\baray{\begin{eqnarray}}
\def\earay{\end{eqnarray}}
\begin{document}
\title{Cosmological Expansion in the Randall-Sundrum Brane World Scenario}

\author{\'Eanna \'E. Flanagan\footnote{eef3@cornell.edu}, S.-H. Henry Tye\footnote{tye@mail.lns.cornell.edu} and Ira Wasserman\footnote{ira@spacenet.tn.cornell.edu}}

\address{
Laboratory for Nuclear Studies and Center for Radiophysics and Space Research\\
Cornell University \\
Ithaca, NY 14853}
\medskip
\date{\today}
\maketitle
\bigskip

\begin{abstract} 
The cosmology of the Randall-Sundrum scenario for a 
positive tension brane in a 5-D Universe with localized 
gravity has been studied previously.  In the radiation-dominated
Universe, it was suggested that there are two solutions for the cosmic
scale factor $a(t)$ : the standard solution $a\sim t^{1/2}$, and a
solution $a\sim t^{1/4}$, which is incompatible with 
standard big bang nucleosynthesis.  
In this note, we reconsider expansion of the Universe in this scenario.
We derive and solve a first order, linear differential equation for $H^2$,
the square of the expansion rate of the Universe, as a function of
$a$.  The differences between 
our equation for $H^2$ and the relationship found in standard
cosmology are (i) there is a term proportional to density squared (a
fact already known), 
which is small when the density is small compared to the brane
tension, and (ii) there is a contribution which acts like a
relativistic fluid.  We show that this second contribution is due to
gravitational degrees of freedom in the bulk.
Thus, we find that there need not be any conflict between cosmology of
the Randall-Sundrum scenario and the standard model of cosmology.  
We discuss how reheating at the end of inflation leads to the correct
relationship between matter density and expansion rate, 
$H^2\to 8\pi G\rho_m/3$, and the conditions that must be met for
the expansion rate of the Universe to be close to its standard model
value around the epoch of cosmological nucleosynthesis.

\end{abstract}

\vfill\eject

\section{Introduction}
\label{sec:intro}

Recently Randall and Sundrum \cite{RS} presented a new static 
solution to the $5$-D (classical) Einstein equations in which 
spacetime is flat on a 3-brane with positive tension provided that the
bulk has an appropriate negative cosmological constant. Even if the fifth
dimension is uncompactified, standard $4$-D gravity (specifically,
Newton's force law) is reproduced on the brane. In contrast to the 
compactified case \cite{ADD}, this follows because the near-brane geometry
traps the massless graviton. 

To see if such a scenario is viable phenomenologically, one
application to check is the evolution of the early Universe \cite{kaloper,others}.  It was
pointed out \cite{binetruy} that a 5-D Universe with branes
may have a rather unconventional, and 
perhaps unacceptable, cosmology. 
The inclusion of matter inside the brane is suggested naturally by the 
brane world picture. So it is important to see if the standard expanding 
Universe can be recovered by extending the static solution to a 
time-dependent one when matter/radiation inside the brane is included. 
As shown in Ref\cite{csaki}, the standard {\it matter}-dominated expanding 
Universe is recovered for large enough brane tension. However, Ref\cite{csaki}
proposed {\it two} possible behaviors for the cosmic
scale factor, $a(t)$, during the {\it radiation}-dominated Universe:
(i)  $a(t) \sim t^{1/4}$ (first found in Ref\cite{binetruy}), or
(ii) $a(t) \sim t^{1/2}$ (found in Ref\cite{csaki}). That there might
be two different powerlaw solutions is not surprising,
since a non-linear second order differential 
equation may have more than one such solution, and initial conditions will 
determine the precise time evolution of $a(t)$ (which  may not be exactly
a powerlaw, but might be close to either of the candidate powerlaws at
different times). On the other hand,
to match the known observations of the expanding Universe, at least back 
to the time of electron-positron annihilation and nucleosynthesis, the
expansion rate of the Universe should be approximately $H^2=8\pi G\rho/3$,
its value in standard Big Bang cosmology. Actually this requirement is
stronger than demanding $a(t)\sim t^{1/2}$: agreement with the well-established
picture of light element synthesis in the early Universe
\cite{peebles} constrains $H$
as a function of temperature $T$ for $T\sim$ 100 keV -- a few MeV.

In this note, we clarify the situation with respect to the cosmology
of the Randall-Sundrum scenario.  First, we elucidate why the
evolution equation that one obtains for the scale factor [Eq.\
(\ref{basiceq}) below] is second order in time rather than first order
in time as in standard cosmology.  The reason is that a piece of the
gravitational dynamics in the bulk is coupled to the brane dynamics,
and thus there is an extra free parameter in the cosmological
equations, describing the amount of effective 4-D energy density due
to the 5-D gravitational degrees of freedom.  The most general bulk
metric compatible with homogeneity and isotropy is a black hole
anti-deSitter metric \cite{schmidt}, and the free parameter is
determined by mass parameter of this bulk solution.  This conclusion
has also been reached by Kraus \cite{Kraus}.

Second, we derive and solve a {\it first}-order 
differential equation for $H^2$ as a function of $\log a$.  This
equation shows that the effect of bulk gravitational degrees of
freedom on the 4-D dynamics is to add a new term that decays with
expansion just like a relativistic fluid (but which may be positive
or negative). This new contribution to $H^2$ is degenerate with
the contributions from relativistic particles such as photons and
neutrinos in the early Universe. Thus, if significant, it would
alter the relationship between $H$ and temperature $T$ of the early
Universe, even though it would still allow $a(t)\sim t^{1/2}$.
Since it is the dependence of $H$ on $T$ that is constrained by
comparing observed light element abundances with the theory of
cosmological nucleosynthesis\cite{peebles,schramm}, 
the new terms due to bulk gravitational
degrees of freedom must be small enough to preserve the concordance
between theory and observation. We show that if we start with
an inflationary epoch \cite{guth,kaloper,albrecht}, and follow the Universe
as it evolves into the radiation-dominated phase, it is possible
(even likely) that the new term becomes insignificant by the time
the Universe reheats at the end of inflation. Our final result for
$H^2$ shows that standard cosmology can be recovered at late
times in the Randall-Sundrum scenario in both the radiation and matter
dominated eras.


\section{Derivation of the brane dynamics}
\label{sec:power}

In this section we derive the equations governing the dynamics of
matter and geometry on the brane, using a local expansion of the
metric near the brane.  We consider a 5-D spacetime with coordinates
$x^A = 
(x^0,x^1,x^2,x^3,x^4) = (t,x^1,x^2,x^3,y)$, and we assume that there
is a single brane located at $y=0$.  We assume the following form for
the 5-D metric \footnote{Bin\'etruy et al. \cite{binetruy} used the 
metric $$ds^2=-n^2(y,t)dt^2+b^2(y,t)dy^2+a^2(y,t)
\delta_{ij}dx^idx^j,$$
but did not employ the freedom to redefine the coordinates $y$ and $t$
to put the metric in the simpler form (\ref{metric0}) used here.  
}
\be
ds^2=\exp[2\beta(y,t)](-dt^2+dy^2)
+\exp[2\alpha(y,t)]\delta_{ij}dx^idx^j.
\label{metric0}
\ee
Thus, the metric only depends on $t$ and $y$, and is flat in ordinary
3-D space (labeled by Latin indices which run over 1,2,3).  For
simplicity, we also restrict attention to the 
case where $\alpha$ and $\beta$ are even functions of $y$.  
For this metric the non-zero components of the Einstein tensor, as
shown by Bin\'etruy et al. \cite{binetruy}, are 
\be
G_{00} = 3 \left[ {\dot \alpha}^2 + {\dot \alpha} {\dot \beta} -
\alpha^{\prime\prime} - 2 \alpha^{\prime\,2} + \alpha^\prime
\beta^\prime \right],
\label{G00}
\ee
\be
G_{ij} = \delta_{ij} e^{2 (\alpha - \beta) } \left[ - 2 {\ddot 
\alpha} - 3 {\dot \alpha}^2 - { \ddot \beta} 
+ 2 \alpha^{\prime\prime} + 3 \alpha^{\prime\,2} 
+\beta^{\prime\prime} \right],
\ee
\be 
G_{44} = G_{yy} = 3 \left[ - {\ddot \alpha} - 2 {\dot \alpha}^2 +
{\dot \alpha} {\dot \beta} + \alpha^{\prime\,2} + \alpha^\prime
\beta^\prime \right],
\ee
and
\be
G_{04} = 3 \left[\beta^\prime {\dot \alpha} + \alpha^\prime {\dot \beta} -
{\dot \alpha}^\prime - {\dot \alpha} \alpha^\prime \right],
\label{G04}
\ee
where dots denote derivatives with respect to $t$ and primes with
respect to $y$.

We assume
that $\alpha$ and $\beta$ are smooth functions of $|y|$ and of $t$,
i.e., $\alpha(y,t) = {\hat \alpha}(|y|,t)$ and $\beta(y,t) = {\hat
\beta}(|y|,t)$, where the functions ${\hat \alpha}(\xi,t)$ and ${\hat
\beta}(\xi,t)$ are smooth in a neighborhood of $\xi=0$.
Then the derivative $\partial \alpha / \partial y$ will generically be
discontinuous across the brane, as in Ref.\ \cite{RS}.  We define
\be
\alpha_1(t) = {\rm Lim}_{y\to0^+} {\partial \alpha(y,t) \over \partial y} = 
\left. {\partial {\hat \alpha}(\xi,t) \over \partial \xi}\right|_{\xi=0}.
\label{alpha1def}
\ee
and 
\be
\beta_1(t) = {\rm Lim}_{y\to0^+} {\partial \beta(y,t) \over \partial y} = 
\left. {\partial {\hat \beta}(\xi,t) \over \partial \xi}\right|_{\xi=0}.
\label{beta1def}
\ee
It follows that
\be
\alpha_{,yy}(y,t) = {\hat \alpha}_{,\xi\xi}(|y|,t) + \alpha_1(t)
\delta(y).
\label{decompos}
\ee
together with a similar equation for $\beta_{,yy}$.

We now substitute the relation (\ref{decompos}) into the Einstein
tensor components (\ref{G00})---(\ref{G04}), and insert into the 5-D
Einstein equations $G_{AB} = \kappa^2 T_{AB}$ where the energy
momentum tensor\footnote{The representation of the brane
stress-energy tensor as a 
$\delta$-function in $y$ involves the tacit assumption
that the thickness of the brane is smaller than $\vert\bo\vert^{-1}$
or $\vert\ao\vert^{-1}$.} is
\be
T^A_{\ \ B}={\rm diag}[(-\rho,p,p,p,0)\delta(y)\exp(-\beta)
+(\Lb,\Lb,\Lb,\Lb,\Lb)].
\ee
Here $\rho$ and $p$ are the density and pressure of the matter on the
brane, and $\Lb>0$ is proportional to minus the bulk cosmological
constant \footnote{The negative bulk cosmological constant ensures
that one solution of the bulk equations is anti-deSitter spacetime \cite{RS}.}.
The result is, first, two equations obtained by equating the
coefficients of the 
$\delta(y)$ in the $G_{00}$ and $G_{ij}$ equations:
\baray
\label{eq:eqoa}
\ao&=&-{\kappa^2\rho\over 6}\ \exp[ \beta(0,t)]\\
\bo&=&{\kappa^2(2\rho+3p)\over 6} \, \exp[\beta(0,t)].
\label{eq:eqo}
\earay
Second, there are the smooth pieces of the equations.  Since we can
restrict attention to $y>0$ for these pieces of the equations (by the
evenness assumption), we can
drop the distinction between ${\hat \alpha}$ and $\alpha$, and between
$\xi = |y|$ and $y$, so we replace terms like ${\hat
\alpha}_{,\xi\xi}(|y|,t)$ [cf.\ Eq.\ (\ref{decompos}) above] by
$\alpha_{,yy}(y,t)$.  The resulting equations are, for $y > 0$,
\be
{\dot \alpha}^2 + {\dot \alpha} {\dot \beta} -
\alpha^{\prime\prime} - 2 \alpha^{\prime\,2} + \alpha^\prime
\beta^\prime = - {\kappa^2 \over 3} \Lambda_b e^{2 \beta},
\label{EE00}
\ee
\be
- 2 {\ddot \alpha} - 3 {\dot \alpha}^2 - { \ddot \beta} 
+ 2 \alpha^{\prime\prime} + 3 \alpha^{\prime\,2} 
+\beta^{\prime\prime} = \kappa^2 \Lambda_b e^{2 \beta},
\ee
\be 
 - {\ddot \alpha} - 2 {\dot \alpha}^2 +
{\dot \alpha} {\dot \beta} + \alpha^{\prime\,2} + \alpha^\prime
\beta^\prime = {\kappa^2 \over 3} \Lambda_b e^{2 \beta},
\ee
and
\be
\beta^\prime {\dot \alpha} + \alpha^\prime {\dot \beta} -
{\dot \alpha}^\prime - {\dot \alpha} \alpha^\prime =0.
\label{EE04}
\ee

\subsection{Power Series Expansion for the Metric Near the Brane}
\label{sec:powerseries}

The equations (\ref{eq:eqoa}) -- (\ref{EE04}) define the coupled
dynamics of the brane and bulk degrees of freedom.  In this section we
derive a description of the dynamics of the brane by itself, by using
a power series expansion for the metric near the brane.

We assume that, near $y=0$,
\baray
\alpha(y,t)&=&\az(t)+\ao(t)\vert y\vert+{1\over 2}\a2(t)y^2
+\cdots\nonumber\\
\beta(y,t)&=&\bz(t)+\bo(t)\vert y\vert+{1\over 2}\b2(t)y^2
+\cdots.
\earay
This assumption is compatible with our definitions (\ref{alpha1def})
and (\ref{beta1def}) of $\alpha_1$ and $\beta_1$.  If we insert this
expansion into Einstein's equations we obtain once again the relations
(\ref{eq:eqoa}) and (\ref{eq:eqo}) on the brane.  Equating powers of
the smooth pieces (\ref{EE00}) -- (\ref{EE04}) of the equations of
motion yields, first, from the piece of the
$G_{04}$ equation (\ref{EE04}) that is $\propto y^0$,
\be
\dot\rho+3\dot\az(\rho+p)=0,
\label{eq:eqcons}
\ee
and five others that follow from the remaining terms,
\be
\dot\az^2+\dot\bz\dot\az-2\ao^2+\ao\bo-\a2=-{\kappa^2\Lb\exp(2\bz)\over 3}
\label{eq:eq1}
\ee
\be
-2\ddot\az-3\dot\az^2-\ddot\bz+3\ao^2+2\a2+\b2=\kappa^2\Lb\exp(2\bz)
\label{eq:eq2}
\ee
\be
-\dot\a2+\a2(\dot\bz-\dot\az)+\b2\dot\az+\bo\dot\ao+\dot\bo\ao
-\dot\ao\ao=0
\label{eq:eq3}
\ee
\be
-\ddot\az-2\dot\az^2+\dot\az\dot\bz+\ao(\ao+\bo)={\kappa^2\Lb\exp(2\bz)\over 3}
\label{eq:eq4}
\ee
\be
\a2(2\ao+\bo)+\ao\b2-\ddot\ao-4\dot\az\dot\ao+\dot\az\dot\bo+\dot\bz\dot\ao
={2\kappa^2\Lb\bo\exp(2\bz)\over 3}.
\label{eq:eq5}
\ee
Equation (\ref{eq:eqcons}) is simply the usual conservation of
energy equation, and its derivation from the $G_{04}$ component
of the 5-D Einstein equations was already given by Bin\'etruy et al.
\cite{binetruy}. Using Eq.\ (\ref{eq:eqo}) in Eq.\ (\ref{eq:eq4})
we find 
\be
-\ddot\az-2\dot\az^2+\dot\az\dot\bz=\biggl({\kappa^2\Lb\over 3}
+{\kappa^4\rho(\rho+3p)\over 36}\biggr)\exp(2\bz);
\ee
defining a new time variable $\that$ by $\dthat=\exp(\bz)dt$, we obtain
the equation already found by Bin\'etruy et al. \cite{binetruy} 
and Csaki et al. \cite{csaki},
\be
-{d^2\az\over\dthat^2}-2\biggl({d\az\over\dthat}\biggr)^2={\kappa^2\Lb\over 3}
+{\kappa^4\rho(\rho+3p)\over 36}.
\label{basiceq}
\ee
Note that $\that$ is just the conventional cosmological proper time,
that is, the proper time as measured by comoving observers on the
brane.  If we assume that 
$\rho=\sigma+\rho_m$ and $p=-\sigma+p_m$, 
where $\sigma$ represents the contribution from the brane
tension, and $\rho_m$ and $p_m$ are the density and pressure
due to the matter, then this equation reduces to
\be
-{d^2\az\over\dthat^2}-2\biggl({d\az\over\dthat}\biggr)^2
={\kappa^2\Lb\over 3}
-{\kappa^4\sigma^2\over 18}
+{\kappa^4\sigma(3p_m-\rho_m)\over 36}
+{\kappa^4\rho_m(\rho_m+3p_m)\over 36}.
\label{eq:eq6}
\ee
We shall find exact solutions of Eq.\ (\ref{eq:eq6}) below.

To complete the solution, note that Eqs. (\ref{eq:eq1}), (\ref{eq:eq2})
and (\ref{eq:eq5}) are algebraic equations for $\a2$ and $\b2$.
Equation (\ref{eq:eq1}) implies that
\baray
\a2&=&\dot\az^2+\dot\az\dot\bz-2\ao^2+\ao\bo+{\kappa^2\Lb\exp(2\bz)\over 3}
\nonumber\\
&=&\dot\az^2+\dot\az\dot\bz+\biggl({\kappa^2\Lb\over 3}
-{\kappa^4\rho(4\rho+3p)\over 36}\biggr)\exp(2\bz),
\earay
where Eq.\ (\ref{eq:eqo}) was used to get the second expression.
Using this result for $\a2$ in Eq.\ (\ref{eq:eq2})
gives
\be
\b2={\kappa^2\Lb\exp(2\bz)\over 3}+\dot\az^2+2\ddot\az+\ddot\bz
+\ao^2-2\ao\bo\nonumber\\
=-3\dot\az^2+\ddot\bz+\biggl({\kappa^4\rho^2\over 12}
-{\kappa^2\Lb\over 3}\biggr)\exp(2\bz),
\ee
where we used eqs. (\ref{eq:eqo}) and (\ref{eq:eq4}) in the
last line. 
It is easy to check that once we have determined $\a2$ and $\b2$,
the remaining equations (\ref{eq:eq3}) and 
(\ref{eq:eq5}) are both satisfied identically. 
Note that the function $\bz(t)$ is not determined from this power series
expansion near the brane.   Thus, it is not specified
by physics on the brane itself.  It is determined by the dynamics in
the bulk, but is gauge dependent.

\section{An Equation for $H^2$}
\label{sec:hsq}

We now turn to the task of solving the second order dynamical equation
(\ref{eq:eq6}) for $\alpha_0(t)$.
The easiest way to do this is to rewrite it as a first order
equation for $H^2=(d\az/\dthat)^2$ as a function of $\az$.  This
rewriting may be accomplished by noting that
\be
{d^2\az\over\dthat^2}
={d(d\az/d\that)\over d\az}{d\az\over\dthat}=H{dH\over d\az}
={1\over 2}{dH^2\over d\az};
\ee
then Eq.\ (\ref{eq:eq6}) becomes
\baray
-{1\over 2}{dH^2\over d\az}-2H^2&=&
-{\exp(-4\az)\over 2}{d[H^2\exp(4\az)]\over d\az}
\nonumber\\
&=&{\kappa^2\Lb\over 3}
-{\kappa^4\sigma^2\over 18}
+{\kappa^4\sigma(3p_m-\rho_m)\over 36}
+{\kappa^4\rho_m(\rho_m+3p_m)\over 36},
\label{eq:binetruy}
\earay
or,
\be
{d[H^2\exp(4\az)]\over d\az}
=\exp(4\az)\biggl[{\kappa^4\sigma^2\over 9}-{2\kappa^2\Lb\over 3}
+{\kappa^4\sigma(\rho_m-3p_m)\over 18}
-{\kappa^4\rho_m(\rho_m+3p_m)\over 18}\biggr].
\label{eq:eq7}
\ee
Note that Eq.\ (\ref{eq:eq7}) represents a linear, {\it first}-order
equation for $H^2$, if we regard the right hand side as a source term
that is (implicitly) a function of $\alpha_0$.
Implicit in our change of variables from $t$ to $\az$ is
the restriction to phases of the evolution of the Universe in which
the two variables are related to one another monotonically. For
oscillating Universes, or Universes in which the scale factor $a(
t) = \exp[\alpha_0(t)]$ may have a 
contracting phase as well as one or more expanding ones, we can derive
Eq.\ (\ref{eq:eq7}) for each of these phases separately.

To solve Eq.\ (\ref{eq:eq7}) generally, let us first rewrite the equation of
energy conservation, Eq.\ (\ref{eq:eqcons}), as
\be
{d\rho_m\over d\az}+3(\rho_m+p_m)=0;
\label{eq:eqcons2}
\ee
this equation implies
\be
p_m=-\rho_m-{1\over 3}{d\rho_m\over d\az}.
\ee
Substituting this result for $p_m$ into equation Eq.\ (\ref{eq:eq7})
yields
\baray
{d[H^2\exp(4\az)]\over d\az}
&=&\exp(4\az)\biggl[{\kappa^4\sigma^2\over 9}-{2\kappa^2\Lb\over 3}
+{\kappa^4\sigma\over 18}\biggl(4\rho_m+{d\rho_m\over d\az}\biggr)
+{\kappa^4\over 36}\biggl(4\rho_m^2+{d\rho_m^2\over d\az}\biggr)
\nonumber\\
&=&{d\over d\az}\biggl[\exp(4\az)
\biggl({\kappa^4\sigma^2\over 36}-{\kappa^2\Lb\over 6}
+{\kappa^4\sigma\rho_m\over 18}
+{\kappa^4\rho_m^2\over 36}\biggr)\biggr];
\earay
we can easily read off the solution
\be
H^2={\kappa^4\sigma^2\over 36}-{\kappa^2\Lb\over 6} 
+{\kappa^4\sigma\rho_m\over 18} 
+{\kappa^4\rho_m^2\over 36}
+K\exp(-4\az),
\label{eq:hsqeq}
\ee
where $K$ is a constant of integration, and may be positive or negative.
In the case of the Randall-Sundrum static solution \cite{RS}, we have
$\kappa^2 \sigma^2 = 6 \Lambda_b$, $\kappa^4 \sigma = 48 \pi G$, and
$\rho_m = K = 0$, giving $H = 0$ and $a(t) = \exp[\alpha_0(t)]=$ constant.
An equation similar to Eq.\ (\ref{eq:hsqeq}) has been derived by Kraus
using a different approach [Eq.\ (27) of Ref.\ \cite{Kraus}].

Just as in standard cosmology based on 4-D general relativity,
the evolution of the Universe can be reduced to the solution of an
equation for $H^2$ and an equation for energy conservation. However,
the equation for $H^2$ has a different structure than in standard
cosmology for three reasons. First, there are two terms that
result from the brane tension and the negative cosmological constant
in the bulk; in the Randall-Sundrum scenario, these terms may be chosen
to cancel exactly. Second, in addition to the normal term proportional to the
matter density $\rho_m$, there is a term that is proportional to
$\rho_m^2$. This high-density ``correction'' term, which is reminiscent 
of the small-scale deviation from Newton's law found in this 
scenario, becomes unimportant once $\rho_m\ll\sigma$. Third,
there is the term $K \exp(-4 \alpha_0)$
that arises purely from initial conditions. This
is a qualitatively new feature of the Randall-Sundrum scenario.
In cosmology based on 4-D general relativity, $H^2$
is completely determined by the energy density of the Universe
(presuming a spatially flat model, as was done above).  But in
the reduction from five dimensions to four dimensions done here, we find
that $H^2$ can be specified freely at some initial time.  The
additional term that results decays exactly as any relativistic
matter density would (although it need not be positive).  

The physical interpretation of the $K \exp(-4 \alpha_0)$ is as
follows.  The most general solution of the bulk equations of motion
(\ref{EE00})---(\ref{EE04}) is the black hole solution \cite{Kraus}
\be
ds^2 = -w(r) d{\bar t}^2 + w(r)^{-1} dr^2 + r^2 \delta_{ij} dx^i dx^j,
\label{bulksoln}
\ee
where
\be
w(r) = {1 \over 6} \Lambda_b r^2 + {m \over r^2}.
\ee
The parameter $K$ is determined by the mass
parameter $m$ of the solution (\ref{bulksoln}); see Eq.\ (27) of
Ref.\ \cite{Kraus}.  Positive $K$ corresponds to $m>0$ and negative
$K$ to $m<0$. Although the solution (\ref{bulksoln})
is static, the motion of the brane through the bulk need not respect
the ${\bar t}$-time translation symmetry, and thus there is a
contribution to the 
effective 4-D energy density which varies with brane proper time
\cite{Kraus}. 

In specific cosmological scenarios, the additional term can be
important or negligible, depending on the magnitude of $K$.  
More concretely, let us imagine that the Universe begins with an
early, inflationary phase of expansion, during which the energy
density and pressure are dominated by an inflaton field, and that
once the inflaton approaches its potential minimum, its energy
dissipates into relativistic particles. For example, in one
model for inflation and reheating (e.g. \cite{albrecht})
\baray
\rho_m&=&{1\over 2}\biggl({d\phi\over\dthat}\biggr)^2
+V(\phi)+\rho_r\nonumber\\
p_m&=&{1\over 2}\biggl({d\phi\over\dthat}\biggr)^2
-V(\phi)+{\rho_r\over 3},
\label{eq:rhopm}
\earay
where the inflaton field $\phi$ obeys the equation
\be
{d^2\phi\over\dthat^2}+(3H+\Gamma_\phi){d\phi\over\dthat}
+{dV(\phi)\over d\phi}=0,
\label{eq:infld}
\ee
and the radiation energy density $\rho_r$
is determined from
\be
{d\rho_r\over\dthat}+4H\rho_r=\Gamma_\phi
\biggl({d\phi\over\dthat}\biggr)^2.
\label{eq:rhor}
\ee
Here $\Gamma_\phi$ is a decay rate that leads to the
production of relativistic particles, primarily from oscillations
of the inflaton field about its minimum. It is straightforward
to set up and solve these equations along with Eq.\ (\ref{eq:eq6})
numerically, and we have done so. To begin the
integration, we must specify not only initial values of $\phi$,
$d\phi/\dthat$ and $\rho_r$,  but also 
the starting value of $H$. Our (limited) exploration\footnote{We chose
$d\phi/\dthat=0$ 
and $\rho_r=0$ initially in all of our numerical integrations, and also 
focussed on cases where the Randall-Sundrum condition, 
$\kappa^2\sigma^2/6\Lb=1$ is satisfied. A simple effective potential of
the form
$V(\phi)=V_0[\exp(-\epsilon)-\exp(-\sqrt{\epsilon^2+\phi^2/\phi_0^2})]$  
was assumed, with $\epsilon$ chosen to be small but nonzero to guard
against pathologies in $dV(\phi)/d\phi$ near $\phi=0$. Although $V_0/\sigma$
is a parameter that can be chosen arbitrarily, so far we have considered
$V_0/\sigma\leq 1$.} 
of numerical solutions shows that, as expected, Eq.\ (\ref{eq:hsqeq}) is
satisfied (to the accuracy of our numerical integrations)
during both the inflationary epoch and the radiation-dominated
era that follows reheating, even though our solutions were
based on Eq.\ (\ref{eq:eq6}) (with independent variable $\az$
instead of $t$), not Eq.\ (\ref{eq:hsqeq}).

Using Eq.\ (\ref{eq:hsqeq}) along with eqs. (\ref{eq:rhopm}), (\ref{eq:infld})
and (\ref{eq:rhor}), we can gain insight into how inflation might
be affected by the new features of our relationship for $H^2$. We assume
that the Randall-Sundrum condition, $\kappa^2\sigma^2/6\Lb=1$,
holds. Let us first
consider the effect of choosing the initial value of $H$, which translates
(with suitable definition of $\az=0$) into choosing a value of $K$.
To keep matters as simple as possible, let us assume that $V_0$, the
value of $V(\phi)$ in its flat portion (where we presume the inflaton
starts), is smaller than $\sigma$, so we can neglect the nonlinear term
in Eq.\ (\ref{eq:hsqeq}) for $H^2$. Then if the initial value of $K$ 
is positive and $\sim
8\pi GV_0/3$, the starting value of $H^2$ in inflation theory
based on 4-D general relativity, deviations in $H^2$ from its standard
cosmological value will damp away exponentially as the inflaton rolls
toward its minimum. Given enough expansion, $K\exp(-4\az)$ becomes
negligible compared with $8\pi GV_0/3$
by the end of inflation, and reheating results in a radiation density
that dominates the $K\exp(-4\az)$ term in Eq.\ (\ref{eq:hsqeq})
thereafter. If the initial value of $H$ is far larger than in
standard cosmology, the inflaton hardly rolls at all until the Universe
expands enough 
that $K\exp(-4\az)$ becomes comparable to $8\pi GV_0/3$; the subsequent
evolution is identical to what would transpire if we had started
with $H^2\sim 8\pi GV_0/3$. This scenario can be thought of as including
a pre-inflation, ``radiation'' dominated era, followed by conventional
inflation and reheating.
If the initial value of $H$ is well below its standard value
initially, which means that $K<0$ (but not so small that $H^2<0$), then 
the total number of e-foldings during inflation is diminished,
but can still be large enough that $K\exp(-4\az)$ becomes
unimportant by the time reheating occurs. Excluding negative
$H^2$ amounts to limiting $-K=\vert K\vert$ to values smaller than
$\kappa^4V_0(\sigma+V_0/2)/18$ initially (at some small but nonzero
starting value of $a(t)$; spatially flat models with $K<0$ have no Big Bang).
This might not be unnatural, as we
might expect $V_0\sim\sigma\sim\vert K\vert/\kappa^4\sigma$.

The nonlinear term in Eq.\ (\ref{eq:hsqeq}) can be important if $V_0
\simgreat\sigma$. In that case, the Universe expands faster than in
standard cosmology, but $H$ is still time-independent while $\phi$
is rolling down the flat part of $V(\phi)$, so much of the standard
picture of inflation can be carried over intact. In particular, as
discussed in the previous paragraph, it is possible that 
the Universe expands enough during inflation to render $K\exp(-4\az)$
negligible before reheating.
The number of e-foldings of $a(\that)$ between the time when the
presently observable  
portion of the Universe crossed the horizon and the end of inflation
is $N_H\sim HT_0/V_0^{1/4}H_0$, where $T_0$ and $H_0$ are the present
temperature and Hubble parameter. Since $N_H\propto H$,
larger $H$ means more e-foldings, and, generically, larger primordial density 
fluctuations, which could be problematic.
When reheating is complete, it is still possible that $\rho_r>\sigma$
for awhile, until expansion can reverse the inequality. 
During the phase in which $H^2 \approx \kappa^4 \rho_m^2 / 36$, the
scale factor behaves as $a(t) = \exp[\alpha_0(t)] \sim t^{1/4}$ as
found in Refs.\ \cite{binetruy,csaki}, but once $H^2 \to \kappa^4
\sigma \rho_m / 18$, the scale factor behaves as $a(t) \sim t^{1/2}$.
As long as 
$\sigma$ is large compared with $\sim$(MeV)$^4$, the nonlinear term
in Eq.\ (\ref{eq:hsqeq}) has no effect on cosmological nucleosynthesis.

In the scenario outlined above, inflation plays a central role in
guaranteeing that the evolution of the Universe tends toward the
standard cosmological model by the nucleosynthesis epoch,
at $T\sim$ 100 keV -- a few MeV. It is not sufficient to require that
$a(t)\to t^{1/2}$ at late times to obtain agreement with standard
model predictions for the synthesis of the light elements, since
this solution is obtained for $K\neq 0$, even if 
$K\exp(-4\az)>\kappa^4\sigma V_0/18$, and $H^2\approx
K\exp(-4\az)$. The outcome of cosmological nucleosynthesis
depends on the competition between $H$ and various
particle reaction rates, such as those that create and destroy
protons and neutrons; these reaction rates can be computed as
functions of $T$ and the baryon density of the Universe
\cite{peebles,schramm}.  The comparison 
of the observed light element abundances with the theoretical
calculations constrains the ratio of $\rho_m$ to its standard model
value at $T\sim$ 100 keV -- 1 MeV, which includes contributions
from photons, three neutrino flavors, and, at $T\simgreat m_e$,
electrons and positrons to be very close to one\cite{schramm}. 
Thus, there is little room for additional relativistic species,
such as $K\exp(-4\az)$ for positive $K$. If $K>0$, it
is crucial that $K/(a(t)T)^4$ become negligible well before
nucleosynthesis occurs, as it is (virtually) unaffected by expansion
alone. If $K\exp(-4\az)<0$, it would have to be balanced nearly exactly by
additional particles that are relativistic during the nucleosynthesis
era. While one cannot rule out such a perverse conspiracy {\it a priori},
it would require fine tuning that seems rather unlikely, so even for
negative $K$ it seems that, generically, $K/(a(t)T)^4$ must 
be negligible at $T\sim$ 100 keV -- 1 MeV.
Inflation and reheating provide a natural mechanism for 
decreasing $K/(a(t)T)^4$ to an acceptable level. Indeed, one might
say that, if the Randall-Sundrum brane world scenario proves correct,
inflation is required to preserve the success of the standard Big
Bang cosmology in accounting for the abundances of the light chemical
elements.

\section{CONCLUSIONS AND REMARKS}

We have re-examined the evolution of homogeneous, isotropic
cosmologies in the Randall-Sundrum brane world scenario, building on
the earlier work of Bin\'etruy et.\ al.\ \cite{binetruy} and Csaki
et.\ al.\ \cite{csaki}.  We derived the explicit form of the equations
that governed the dynamics of both the bulk and brane degrees of
freedom [Eqs.\ (\ref{eq:eqoa}), (\ref{eq:eqo}), (\ref{eq:eqcons}) and
(\ref{EE00})---(\ref{EE04}) above].  We derived
a generalization [Eq.\ (\ref{eq:hsqeq}) above] of the standard
relation between the Hubble  
parameter $H$ and energy density $\rho_m$.  This equation contains a
term (an effective 4-D energy density) that acts like a free,
relativistic species, and is the only effect of the 5-D gravitational
degrees of freedom on the bulk dynamics.

The cosmological equation we find for
$H$ contains extra terms compared to 4-D general relativity
(a term proportional to density squared and the term due to 5-D
gravitational degrees of freedom) but, generically, these become
unimportant at 
late times.  In particular, we analyzed how reheating at the end of
inflation 
smoothly matches onto the standard behavior $a(t) \propto t^{1/2}$ of
the radiation dominated epoch, and not the behavior $a(t) \propto
t^{1/4}$ found in Ref.\ \cite{binetruy}. 
We show that inflation can drive the expansion rate of the
Universe toward its standard cosmological value, $H^2=8\pi G\rho_m/3$
after reheating under a variety of conditions. Thus, the Randall-Sundrum
scenario is compatible with standard cosmology.

A recent paper by Shiromizu, Maeda and Sasaki \cite{Sasaki} derives
effective Einstein equations on the brane for a general bulk metric.
Their independent analysis is compatible with our results; they find
corrections to the Einstein equation that scale as density squared, and
also a correction term $E_{\mu\nu}$ due both to bulk gravitational waves
and to non-radiative bulk degrees of freedom.  Bulk
gravitational waves do not arise in our analysis because of
the symmetries we assumed in the starting metric (\ref{metric0}).  On
the other hand, our term $K \exp[-4 \alpha_0(t)]$ in Eq.\
(\ref{eq:hsqeq}) corresponds to the non-radiative piece of the tensor
$E_{\mu\nu}$ of Eq.\ (17) of Ref.\ \cite{Sasaki}.

Finally, we remark that to solve the hierarchy problem, it was
suggested \cite{LR} that we live in a probe brane (or ``TeV'' brane),
which is at a distance $y_0$ away from the Planck brane, where the
graviton is trapped. In this case, the tension of the ``TeV'' brane is
substantially  
smaller than that of the Planck brane. In the cosmological setting, 
depending on the details, the $\rho_m^2$ term contribution to $H^2$
may no longer be negligible on the ``TeV'' brane, and the standard 
cosmology will be modified accordingly.  It will be interesting to 
see if cosmological constraints can be consistent with this
solution to the hierarchy problem.

\acknowledgments

We thank Gary Shiu for helpful discussions.  We are grateful to
P. Kraus for pointing out an error in interpretation in an earlier
version of this paper.  \'E.F.\ was supported in part by NSF grant PHY
9722189 and by the Alfred P. Sloan foundation.  The research of
S.-H.H.T. is partially supported by the National Science Foundation.
I.W. gratefully acknowledges support from NASA.

\end{document}